\documentclass[preprint,preprintnumbers,amsmath,amssymb,superscriptaddress,endfloats]{revtex4}\newif\ifprepring\let\ifpreprint\iftrue

\usepackage{graphicx}
\usepackage{dcolumn}
\usepackage{bm}

\newcommand{\etal}{{\it et al}}
\newcommand{\affCrystMaterSci}{Department of Crystalline Materials Science, %
  Nagoya University, %
  Chikusa-ku, Nagoya 464-8603, Japan}
\newcommand{\affVBL}{Venture Business Laboratory (VBL), %
  Nagoya University, %
  Chikusa-ku, Nagoya 464-8603, Japan}
\newcommand{\affTRIP}{TRIP, Japan Science and Technology Agency (JST), %
  Tokyo 102-0075, Japan}

\begin{document}

\title{Epitaxial Growth of NdFeAsO Thin Films by Molecular Beam Epitaxy}

\affiliation{\affCrystMaterSci}
\affiliation{\affVBL}
\affiliation{\affTRIP}
\author{T.\ Kawaguchi}
  \affiliation{\affCrystMaterSci}
  \affiliation{\affTRIP}
\author{H.\ Uemura}
  \affiliation{\affCrystMaterSci}
  \affiliation{\affTRIP}
\author{T.\ Ohno}
  \affiliation{\affCrystMaterSci}
\author{R.\ Watanabe}
  \affiliation{\affCrystMaterSci}
  \affiliation{\affTRIP}
\author{M.\ Tabuchi}
  \affiliation{\affVBL}
  \affiliation{\affTRIP}
\author{T.\ Ujihara}
  \affiliation{\affCrystMaterSci}
  \affiliation{\affTRIP}
\author{K.\ Takenaka}
  \affiliation{\affCrystMaterSci}
  \affiliation{\affTRIP}
\author{Y.\ Takeda}
  \affiliation{\affCrystMaterSci}
  \affiliation{\affTRIP}
\author{H.\ Ikuta}
  \affiliation{\affCrystMaterSci}
  \affiliation{\affTRIP}

\date{July 17, 2009}

\begin{abstract}
Epitaxial films of NdFeAsO were grown on 
GaAs substrates by molecular beam epitaxy (MBE). 
All elements including oxygen were supplied from solid sources
using Knudsen cells. 
The x-ray diffraction pattern of the film prepared with 
the optimum growth condition showed no indication of impurity phases.
Only (00$l$) peaks were observed,
indicating that NdFeAsO was grown with the $c$-axis perpendicular
to the substrate. 
The window of optimum growth condition was very narrow,
but the NdFeAsO phase was grown with a very good reproducibility.
Despite the absence of any appreciable secondary phase,
the resistivity showed an increase with decreasing temperature.
\end{abstract}

\maketitle

The discovery of high temperature superconductivity in F-doped
LaFeAsO \cite{Kamihara} has ignited a great deal of excitement 
and an explosive flow of studies on 
iron-pnictides and related materials. 
The worldwide efforts led to the increase in 
the critical temperature $T_c$ by substituting other 
lanthanide elements for La,
resulting in a thus far record high $T_c$ of 
56 K \cite{XHChen,RenEPL,Kito,RenCPL,CWang},
and to the finding of other superconductors
with related structures \cite{Rotter,Tapp,Hsu}.
High-quality epitaxial films are indispensable both for
exploring the intrinsic properties of these materials 
and for electronic device applications.
Efforts of preparing thin films of the new superconductors
started soon after their discovery \cite{Hiramatsu1111,Backen}.
Thin films with $T_c$ similar to the bulk values
have already been reported for Co-doped $AE$Fe$_2$As$_2$ ($AE$=Sr, Ba) 
\cite{Hiramatsu122,Katase} and 
iron-chalcogens \cite{Han,MJWang,Nie,Bellingeri,Mele}.
However, the thin film preparation of
the $Ln$FeAsO ($Ln$=lanthanide) family,
which exhibits the highest $T_c$ to date 
among the iron-pnictide superconductors,
seems to be more difficult.
Hiramatsu \etal.\ were the first to successfully prepare
epitaxial films of LaFeAsO \cite{Hiramatsu1111},
although no superconductivity was observed. 
Superconducting F-doped LaFeAsO thin films were
reported from another group,
but only after the films were post-annealed
at significantly high temperatures \cite{Backen,Haindl}.
Obviously, there is a large room for improving
the film quality of $Ln$FeAsO.

All the above mentioned thin films were prepared
by pulsed-laser deposition (PLD) on oxide crystalline substrates.
Among the various thin film preparation techniques, 
molecular beam epitaxy (MBE) 
has been proven as a reliable growth method for 
the fabrication of 
high quality films of many different materials.
The flux from each deposition source can be 
independently and precisely controlled,
which is a large advantage for fine tuning the growth condition.
Therefore, we employed the MBE method in this study,
and report on the first successful growth of 
an epitaxial film of the $Ln$FeAsO family by MBE.

NdFeAsO was chosen as the material to be grown because of 
its high potential in terms of $T_c$.
GaAs(001) were used as substrates.
The lattice matching between GaAs and NdFeAsO is very good,
because the lattice constant $a$ of NdFeAsO (0.3963 nm \cite{Kito})
multiplied by $\sqrt{2}$ is close to 
the lattice constant of GaAs (0.5653 nm).
First, an about 300 nm thick GaAs buffer layer was 
grown at 610$^\circ$C on the substrate.
NdFeAsO was then grown at 670$^\circ$C
by supplying all elements from solid sources 
charged in Knudsen cells;
Fe, As, NdF$_3$, and Fe$_2$O$_3$.
Here, Fe$_2$O$_3$ was used as an oxygen source.
A certain amount of flux was observed
when the cell temperature of Fe$_2$O$_3$ was varied between 
500 and 800$^\circ$C,
although the vapor pressure of Fe is very low in this temperature range.
In fact, the Fe content of the film did not 
change appreciably when only the temperature of 
the Fe$_2$O$_3$ cell was altered.
Therefore, we think that Fe$_2$O$_3$ merely supplies oxygen
without increasing the amount of Fe flux 
for the deposition conditions adopted in this study. 
This is reasonable because 
Fe$_2$O$_3$ is expected to reduce to Fe$_3$O$_4$ 
for the present experimental conditions
from a thermodynamic consideration based on 
the Ellingham diagram \cite{Ellingham}.

The Nd and Fe contents of the prepared films were examined 
by electron probe micro-analysis (EPMA) using 
NdFeAs(O,F) powders as a reference, which were 
prepared by the method reported before \cite{Takenaka}.
To determine the thickness, part of the film was removed 
by etching with hydrochloric acid (HCl), 
and the height of the step was measured using an atomic 
force microscope (AFM).
The film thickness thus determined was 
about 30 nm for a film grown for 2 hours. 
Except for this film,
all other films reported below were grown for 1 hour. 
The relation between the vapor pressure and the cell temperature
was determined using an ion gauge beam flux monitor.
Resistivity was measured by a four-probe method.

\begin{figure}
  \includegraphics[width=0.95\columnwidth]{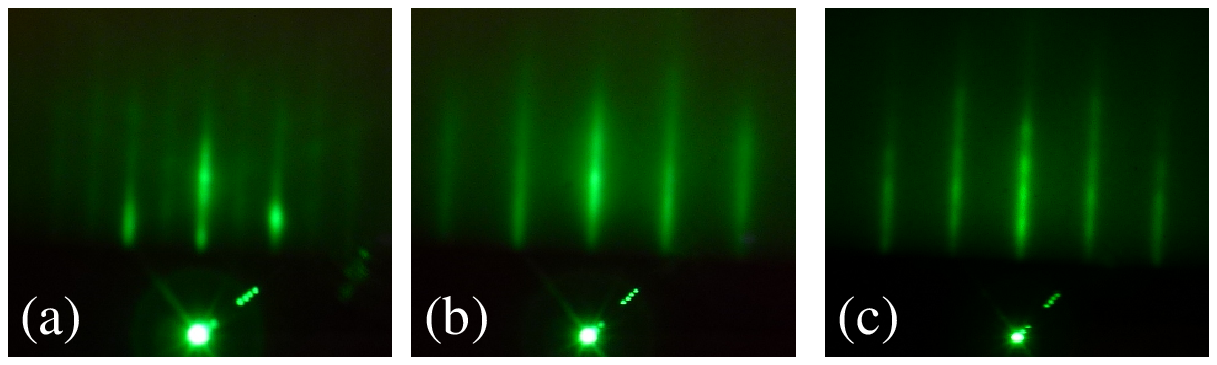}
  \caption{\label{fig:RHEED}
    (Color online) RHEED patterns taken along the GaAs[110] azimuth; 
    (a) after the growth of the buffer layer, 
    (b) 10 min after starting the growth of NdFeAsO,
    and (c) after the film was grown.
  }
\end{figure}

Figure \ref{fig:RHEED} shows 
reflection high-energy electron diffraction (RHEED)
patterns taken along the GaAs[110] azimuth 
during the film growth.
A streaky pattern with a clear (2$\times$4) structure was 
observed after the buffer layer was grown (Fig.\ \ref{fig:RHEED}(a)).
The streaks elongated when the growth of NdFeAsO 
started (Fig.\ \ref{fig:RHEED}(b)),
and were well maintained until the end of the growth (Fig.\ \ref{fig:RHEED}(c)).
These results suggest that the film was grown with a flat surface, 
which is very important for device applications.
The RHEED patterns in the [1$\bar{1}$0] direction 
were very similar to that along [110] azimuth,
implying that the NdFeAsO film was grown 
epitaxially on the GaAs buffer layer.

\begin{figure}[b]
  \ifpreprint\newpage\fi
  \includegraphics[width=0.8\columnwidth]{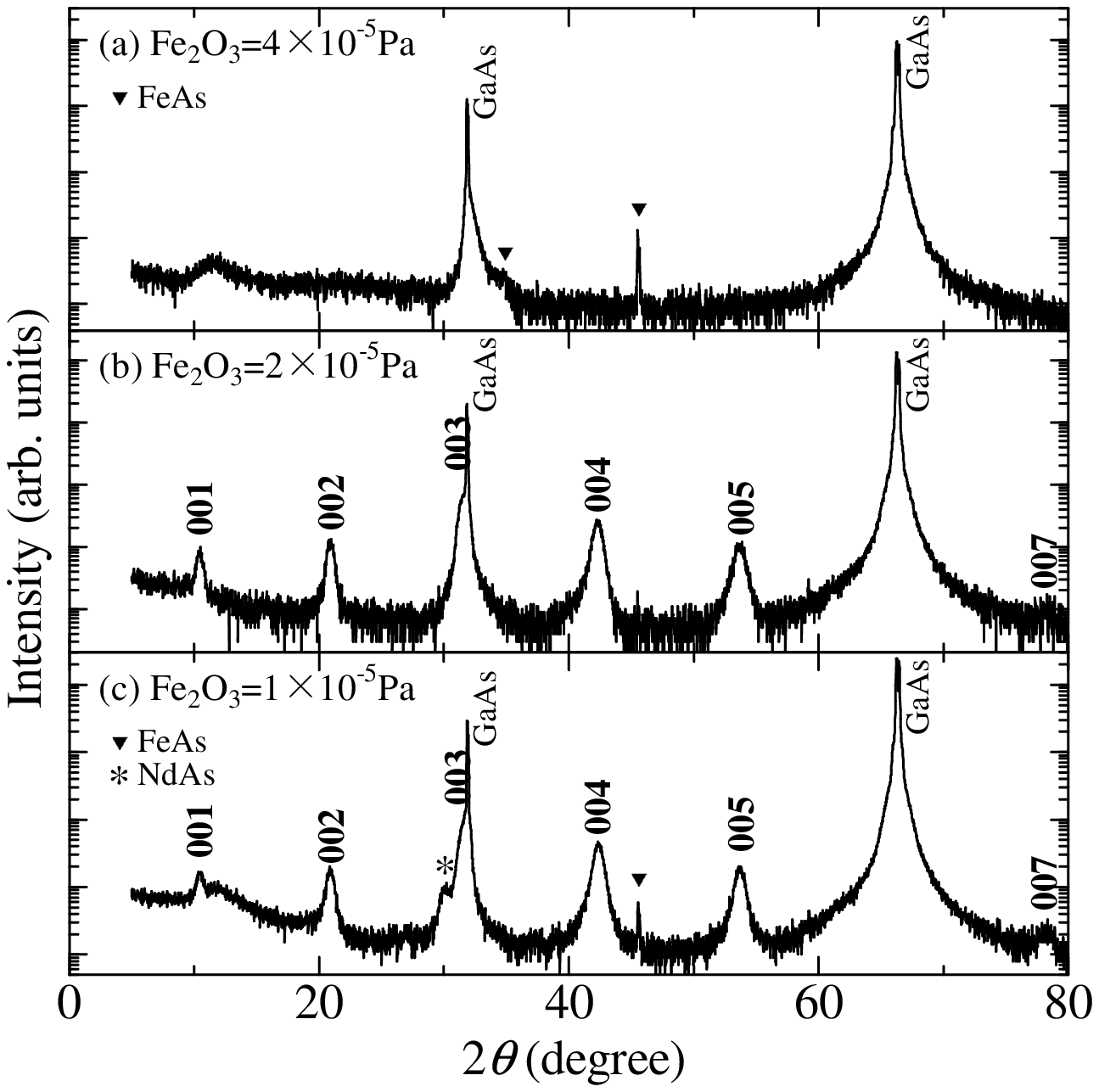}
  \caption{\label{fig:XRDoxygen}
    XRD profiles of films grown with various oxygen fluxes.
    The vapor pressures of Fe, As and NdF$_3$ were 
    1.9$\times$10$^{-6}$ Pa, 1.5$\times$10$^{-3}$ Pa, and
    2.7$\times$10$^{-6}$ Pa, and
    that of oxygen were (a) 4$\times$10$^{-5}$ Pa, 
    (b) 2$\times$10$^{-5}$ Pa, and (c) 1$\times$10$^{-5}$ Pa,
    respectively.
  }
\end{figure}

Since NdFeAsO consists of four elements, there were
many parameters that had to be controlled
to successfully grow an epitaxial film.
We first changed the flux of Nd and Fe and adjusted 
their compositional ratio based on the EPMA data.
After that, the vapor pressures of As and oxygen were changed.
Figure \ref{fig:XRDoxygen} shows x-ray diffraction (XRD) profiles
of films that were grown by changing only the oxygen flux.
The vapor pressures of Fe, As and NdF$_3$ were 
1.9$\times$10$^{-6}$ Pa, 1.5$\times$10$^{-3}$ Pa, and
2.7$\times$10$^{-6}$ Pa, respectively.
Only few peaks that can be assigned to FeAs were observed
beside the substrate peaks for the film grown 
with the largest oxygen flux (Fig.\ \ref{fig:XRDoxygen}(a)).
Peaks that can be indexed as (00$l$) reflections of NdFeAsO 
emerged when the oxygen flux was reduced,
but NdAs was observed as the secondary phase 
when the oxygen flux was too small (Fig.\ \ref{fig:XRDoxygen}(c)).
Only within a small window, 
a single-phased NdFeAsO film was obtained 
as shown in Fig.\ \ref{fig:XRDoxygen}(b).

\begin{figure}
  \ifpreprint\newpage\fi
  \includegraphics[width=0.8\columnwidth]{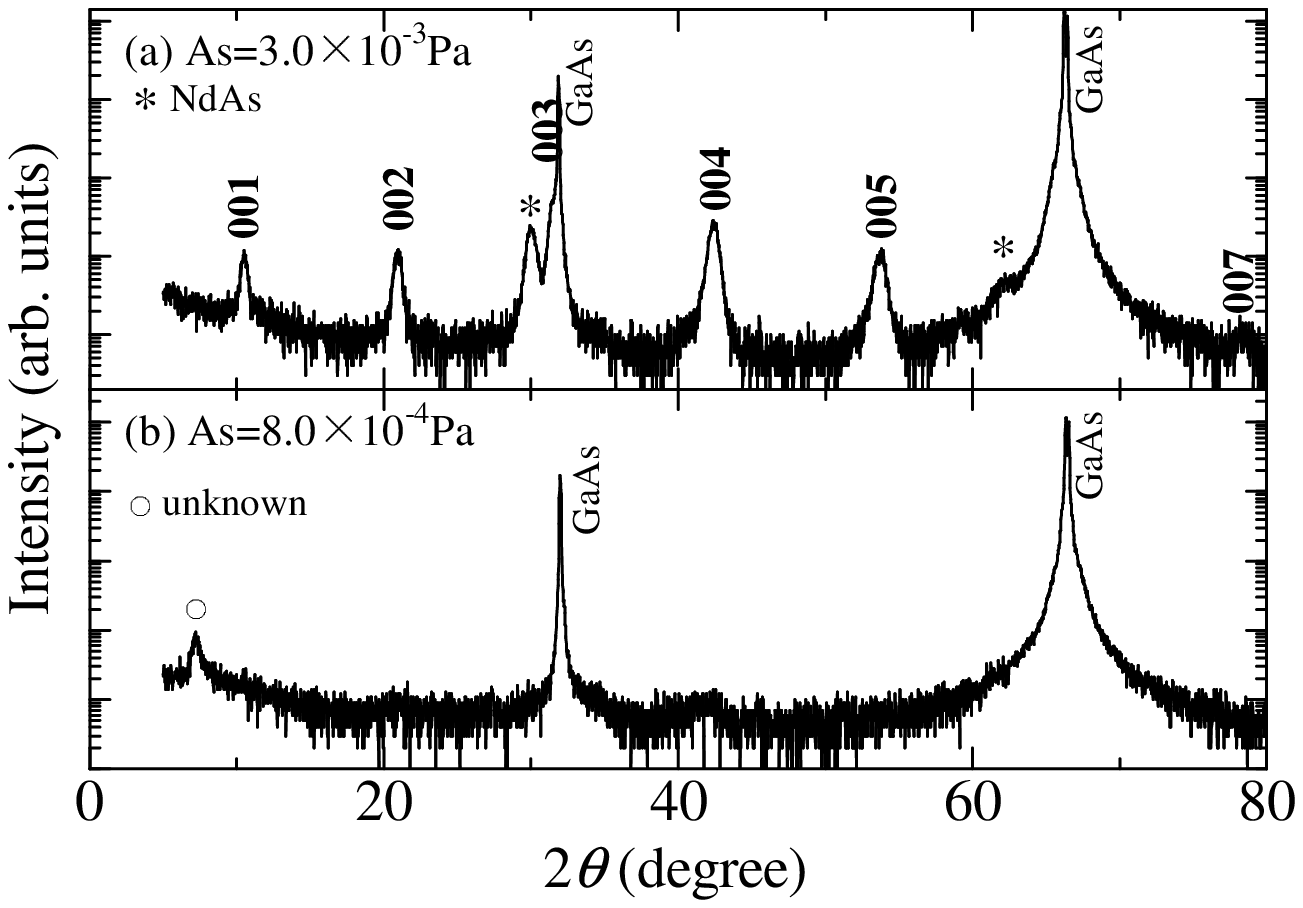}
  \caption{\label{fig:XRDAs}
    XRD profiles of films grown with various As fluxes.
    The vapor pressures of Fe, NdF$_3$, and oxygen were 
    1.9$\times$10$^{-6}$ Pa, 2.7$\times$10$^{-6}$ Pa, and
    2$\times$10$^{-5}$ Pa, and
    that of As were (a) 3.0$\times$10$^{-3}$ Pa, 
    and (b) 8.0$\times$10$^{-4}$ Pa,
    respectively.
  }
\end{figure}

A similar experiment was conducted for the vapor pressure
dependence of As as well.
Figures \ref{fig:XRDAs}(a) and (b) show the XRD profiles
of films that were grown with an As vapor pressure
larger and smaller than that of 
the best film (Fig.\ \ref{fig:XRDoxygen}(b)), respectively.
All the other conditions were the same as that of 
the film of Fig.\ \ref{fig:XRDoxygen}(b).
When As was oversupplied, NdAs was observed 
as an impurity phase (Fig.\ \ref{fig:XRDAs}(a)).
On the other hand, almost no crystalline phase 
was observed for the film grown with the lowest As vapor pressure
(Fig.\ \ref{fig:XRDAs}(b)),
similarly to the film for which oxygen 
was oversupplied (Fig.\ \ref{fig:XRDoxygen}(a)).

\begin{figure}[b]
  \ifpreprint\newpage\fi
  \includegraphics[width=0.7\columnwidth]{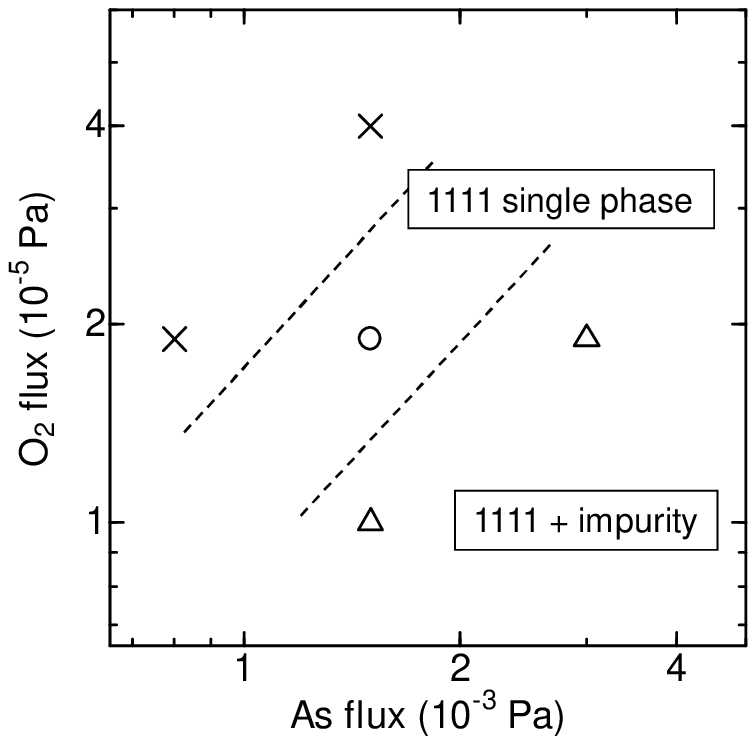}
  \caption{\label{fig:PD}
    The phase diagram for the films prepared at 670$^\circ$C
    with Fe and NdF$_3$ vapor pressures of 
    1.9$\times$10$^{-6}$ Pa and 2.7$\times$10$^{-6}$ Pa, 
    respectively.
    A circle indicates that the resulted film was 
    single-phased NdFeAsO,
    a triangle means that small impurity peaks were observed together
    with the dominant NdFeAsO phase, 
    a cross implies that no NdFeAsO phase was observed.
    The dashed lines indicate only roughly the boundaries.
  }
\end{figure}

Figure \ref{fig:PD} summarizes the results of 
the XRD observations. 
Interestingly, increasing oxygen flux 
had an effect similar to reducing As flux. 
This is probably because arsenic oxides,
which have high vapor pressures, 
are formed when the oxygen flux is large. 
The growth window for a single-phased NdFeAsO film
is very narrow,
as is obvious from the diagram.
Nevertheless, NdFeAsO was always 
the dominant phase when the growth condition was 
close to the optimum one.
This includes films that were grown with 
different vapor pressures of Fe and/or NdF$_3$
(and therefore which cannot be plotted on Fig.\ \ref{fig:PD}).
In other words, NdFeAsO was grown with a very good reproducibility, 
in contrast to the report on PLD preparation 
of LaFeAsO films \cite{Hiramatsu1111}.

The $c$-axis lattice constant 
calculated from the peaks labeled as 002, 004, and 005 
of the optimized film shown in Fig.\ \ref{fig:XRDoxygen}(b)
was 0.857$\pm$0.002 nm, which 
is close to the value reported for bulk samples \cite{Kito}.
On the other hand, the film thickness was estimated to be 
15 nm as mentioned above.
This implies that the growth rate is very low, 
but is consistent to the fact that 
growth of only small single crystals has been reported for 
$Ln$FeAsO so far \cite{Karpinski}.

\begin{figure}
  \ifpreprint\newpage\fi
  \includegraphics[width=0.75\columnwidth]{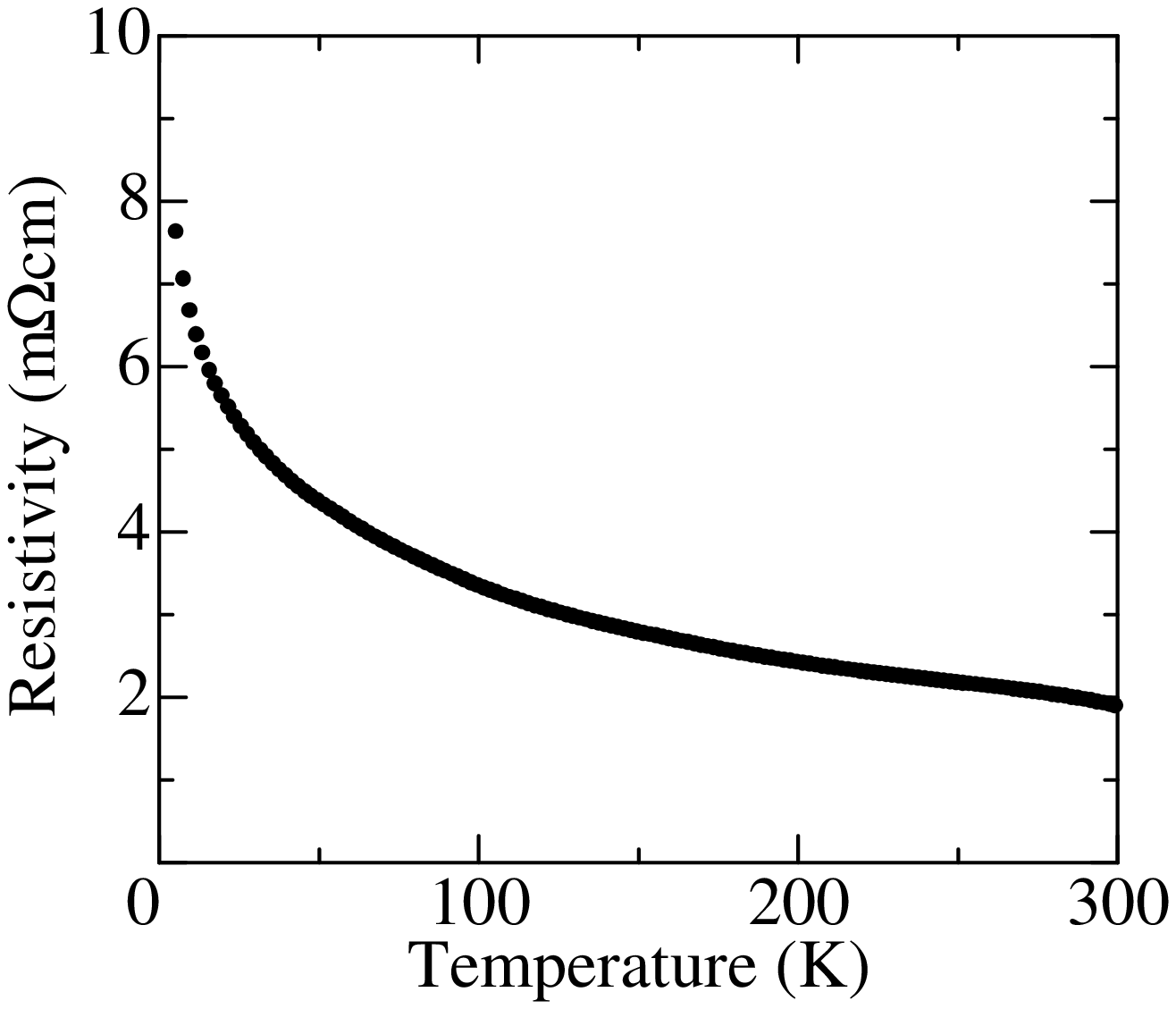}
  \caption{\label{fig:rT}
    Temperature dependence of resistivity of the film 
    of Fig.\ \protect\ref{fig:XRDoxygen}(b).
  }
\end{figure}

Figure \ref{fig:rT} shows the temperature dependence of resistivity 
of the single-phased film of Fig.\ \ref{fig:XRDoxygen}(b).
While a metallic temperature dependence with a drop at about 150 K
due to a structural transition is reported for 
bulk samples \cite{Kamihara},
the resistivity of our film increased with temperature decreasing. 
Such semiconductor-like behaviors were observed
for all the so-far prepared films for which NdFeAsO 
was the dominant phase (e.g.\ Fig.\ \ref{fig:XRDoxygen}(c) and
Fig.\ \ref{fig:XRDAs}(a)).
It is also similar to the temperature dependence of 
the as-grown films prepared by PLD \cite{Hiramatsu1111}.
The reason of the semiconductor-like behavior despite 
the absence of any obvious secondary phase is not clear
but several possibilities may be pointed out. 
For instance, the film might be structurally deteriorated 
because the thickness is only 15 nm, 
the composition might be slightly deviated from the ideal one
and the carrier content is not appropriate,
or Ga might be diffused from the GaAs substrate
because the growth temperature was rather high. 
A further optimization of the growth condition 
is highly desired. 

In conclusion, we have grown NdFeAsO films 
epitaxially on GaAs substrates.
This is the first report of a successful growth of 
$Ln$FeAsO films by MBE. 
All elements including oxygen were supplied from solid sources.
For the growth of the NdFeAsO phase, 
it was important to precisely adjust the flux of 
the constituting elements, 
particularly the balance between oxygen and As. 
Even a small deviation from the optimum condition resulted 
in a formation of impurities, or in the worse case, 
absence of the NdFeAsO phase. 
Despite the narrow window of optimum growth condition, 
a very good reproducibility was obtained.
The RHEED observation suggests that the growth 
proceeded with maintaining a flat surface, 
which is favorable for device applications. 

This work was supported by 
Transformative Research Project on Iron Pnictides (TRIP), JST.


\ifpreprint\newpage\fi

\ifpreprint\newpage\fi

\end{document}